\documentclass[%
 reprint,
superscriptaddress,
 amsmath,amssymb,
 aps,
 twocolumn,
]{revtex4}

\usepackage[colorlinks=true,urlcolor=blue,citecolor=blue,linkcolor=blue]{hyperref}
\usepackage{graphicx}
\usepackage{amsmath}
\usepackage{times}
\usepackage{mathtools}
\usepackage{braket}
\usepackage{rotating}
\usepackage{amssymb}
\newcommand{\neswarrow}{%
        \begin{turn}{45}
                \raisebox{-1ex}{$\leftrightarrow$}
        \end{turn}
}
\newcommand{\nwsearrow}{%
        \begin{turn}{135}
                \raisebox{-0.2ex}{$\leftrightarrow$}
        \end{turn}
}

\begin{document}

\preprint{APS/123-QED}

\title{Direct observation of the particle exchange phase of photons}

\author{Konrad Tschernig$^*$}
\affiliation{Max-Born-Institut f\"ur Nichtlineare Optik und Kurzzeitspektroskopie\\
Max-Born-Stra{\ss}e 2A, 12489 Berlin, Germany}
\affiliation{Institut f\"ur Physik, Humboldt-Universit\"at zu Berlin\\
Newtonstra{\ss}e 15, 12489 Berlin, Germany}
\thanks{These authors contributed equally to this work.}

\author{Chris M\"uller$^*$}
\affiliation{Institut f\"ur Physik, Humboldt-Universit\"at zu Berlin\\
Newtonstra{\ss}e 15, 12489 Berlin, Germany}
\affiliation{IRIS Adlershof, Humboldt-Universit\"at zu Berlin\\
Zum Gro{\ss}en Windkanal 6, 12489 Berlin, Germany}
\thanks{These authors contributed equally to this work.}

\author{Malte Smoor}
\affiliation{IRIS Adlershof, Humboldt-Universit\"at zu Berlin\\
Zum Gro{\ss}en Windkanal 6, 12489 Berlin, Germany}

\author{Tim Kroh}
\affiliation{Institut f\"ur Physik, Humboldt-Universit\"at zu Berlin\\
Newtonstra{\ss}e 15, 12489 Berlin, Germany}
\affiliation{IRIS Adlershof, Humboldt-Universit\"at zu Berlin\\
Zum Gro{\ss}en Windkanal 6, 12489 Berlin, Germany}

\author{Janik Wolters}
\affiliation{Deutsches Zentrum f\"ur Luft- und Raumfahrt e.V. (DLR), Institute of Optical Sensor Systems\\
Rutherfordstra{\ss}e 2, 12489 Berlin, Germany}
\affiliation{Technische Universit\"at Berlin, Institut f\"ur Optik und Atomare Physik\\
Str. des 17. Juni 135, 10623 Berlin, Germany}

\author{Oliver Benson}
\affiliation{Institut f\"ur Physik, Humboldt-Universit\"at zu Berlin\\
Newtonstra{\ss}e 15, 12489 Berlin, Germany}
\affiliation{IRIS Adlershof, Humboldt-Universit\"at zu Berlin\\
Zum Gro{\ss}en Windkanal 6, 12489 Berlin, Germany}

\author{Kurt Busch}
\affiliation{Max-Born-Institut f\"ur Nichtlineare Optik und Kurzzeitspektroskopie\\
Max-Born-Stra{\ss}e 2A, 12489 Berlin, Germany}
\affiliation{Institut f\"ur Physik, Humboldt-Universit\"at zu Berlin\\
Newtonstra{\ss}e 15, 12489 Berlin, Germany}

\author{Armando Pérez-Leija}
\affiliation{Max-Born-Institut f\"ur Nichtlineare Optik und Kurzzeitspektroskopie\\
Max-Born-Stra{\ss}e 2A, 12489 Berlin, Germany}
\affiliation{Institut f\"ur Physik, Humboldt-Universit\"at zu Berlin\\
Newtonstra{\ss}e 15, 12489 Berlin, Germany}

\date{\today}

\begin{abstract}
Quantum theory stipulates that if two particles are identical in all physical aspects, 
the allowed states of the system are either symmetric or antisymmetric with respect to 
permutations of the particle labels. 
Experimentally, the symmetry of the states can be inferred indirectly from the fact that 
neglecting the correct exchange symmetry in the theoretical analysis leads to dramatic 
discrepancies with the observations. The only way to directly unveil the symmetry of the 
states for, say, 
two identical particles is through the interference of the 
original state and the physically permuted one, and measure the phase associated with the 
permutation process, the so-called particle exchange phase. Following this idea, we have 
measured the exchange phase of indistinguishable photons, providing direct evidence of 
the bosonic character of photons. 
\end{abstract}

\maketitle


In three spatial dimensions, quantum physics distinguishes between two fundamental types 
of particles, bosons and fermions \cite{Leinaas1977}. As a result, indistinguishable bosons satisfy 
the Bose-Einstein statistics and indistinguishable fermions obey the Fermi-Dirac statistics. 
Simply put, this means that fermions cannot occupy the same quantum state, as dictated by 
the Pauli exclusion principle \cite{PauliExclusion}, while bosons are allowed to `condensate' 
into the same state \cite{Ketterle}. 
Quite remarkably, these distinctive characteristics are key ingredients that give form to all 
the existing elements and fields as we know them in the universe. \\
From a theoretical perspective, it has been postulated that the correct statistics for bosons 
and fermions can only be observed provided the associated states are symmetric and antisymmetric, 
respectively \cite{SymmetrizationPostulate}. 
This means that, under permutations of the labels of any pair of particles, the allowed states 
of a system of $N$ identical bosons must remain unchanged, 
$\hat{P}_{i,j}\ket{N~\text{bosons}}= \ket{N~\text{bosons}}$, while the states of a system of 
$N$ identical fermions must undergo a sign change, 
$\hat{P}_{i,j}\ket{N~\text{fermions}}= -\ket{N~\text{fermions}}$ \cite{ModernQuantumMechanics}. 
Here, $\hat{P}_{i,j}$ represents the permutation operator that interchanges the labels of the 
$i$'th and $j$'th particle, where $i$ and $j$ are arbitrary.\\ 
Certainly, the (anti-)symmetric nature of the states have very profound implications for 
quantum science and technology \cite{Walmsley1001}. In this respect, perhaps the most prominent example is the so-called Hong-Ou-Mandel effect \cite{HOM}, which sorts two indistinguishable bosons (fermions) into the same (opposite) output channel of a beam splitter \cite{Bocquillon1054}.
Further, it is now recognized that multi-particle states described by (anti-)symmetric states
belong to a very special set of quantum states, referred to as decoherence-free subspaces, which 
are immune to the impact of environmental noise \cite{decoherencefree}. Along similar lines, 
quantum indistinguishability is an inherent control for noise-free entanglement generation 
\cite{Nosrati}, and it represents a source of quantum coherence, even when the particles are 
prepared independently \cite{Castellini}.\\
Beyond explorations of the advantages offered by quantum indistinguishable particles, many 
authors have investigated the validity of the symmetrization postulate in a variety of experiments 
ranging from spectroscopy \cite{Hilborn,Modugno,SpectroscopicTest}, via quantum chemistry 
\cite{Ospelkaus853} to ultracold atoms \cite{Levin}.  
Yet, in all those experiments the postulate has been demonstrated indirectly, e.g. by examining 
the absence of particular states which are forbidden by the postulate 
\cite{Ramberg,Angelis,Modugno,DeMille}.  
In physical terms, the (anti-)symmetrization of the states implies the existence of a definite 
relative phase between the constituent states \cite{Mirman}. 
This concept is more intuitively illustrated for the particular case of two indistinguishable 
particles occupying two different spatial modes $x$ and $y$ with equal probability amplitude 
$1/\sqrt{2}$. 
Thus, the only possibility that satisfies the symmetrization postulate is 
\begin{align}
\ket{\Psi}=\frac{1}{\sqrt{2}}\left(\ket{x}\ket{y}+e^{i\phi_{x}}\ket{y}\ket{x}\right),
\label{eq:WF}
\end{align}
with $\phi_{x}=0$ for bosons and $\phi_{x}=\pi$ for fermions. The argument $\phi_{x}$ is termed 
the \emph{particle exchange phase} (EP) and, in principle, it is amenable to direct measurement 
\cite{Landshoff}. Clearly, such a measurement of the EP would reflect directly the fundamental 
physical symmetry of the associated states \cite{Mirman}.\\
Experimentally, the only way to measure phase differences is via interferometry. That is, to 
observe the EP we have to superpose a reference state $\ket{x}\ket{y}$, with its physically 
permuted version $e^{i\varphi}\ket{y}\ket{x}$. 
However, care has to be exercised: Exchanging the position, or other physical properties, of 
two identical particles yields a final state exhibiting a phase factor ${\varphi}$ composed 
not only of the EP but it also contains an extra phase shift $\phi_{g}$ whose origin is purely 
geometric \cite{Peres}. Essentially, the geometric phase is an observable phase accumulation 
that is acquired by the states of physical systems whose dynamic is cyclic, i.e., when the 
states return to their initial configuration 
\cite{AharonovAnandan,Kai}.
In other words, when considering the actual experimental implementation of exchanging the
states of two identical particles one cannot ignore the geometric phase associated with 
this exchange.\\

In this work, we report on the observation of the particle exchange phase for indistinguishable 
photons. For our experiments, we have implemented an interferometer that superposes a reference 
state of two indistinguishable photons occupying two spatial modes with its physically permuted 
version \cite{CFRoos}. Our direct measurements yield $\phi_x = (-0.04 \pm 0.07) \text{ rad}$, 
which imposes less stringent bounds on a possibly anomalous photon exchange phase than previous indirect measurements as reported in 
\cite{SpectroscopicTest,ComptonTest}.
Concurrently, our observations reveal a geometric phase $\phi_g=\pi$, which is expected from a 
swap operation on identical particles \cite{GeometricPhasevanEnk}.\\

\begin{figure}
    \centering
    \includegraphics[width=\linewidth]{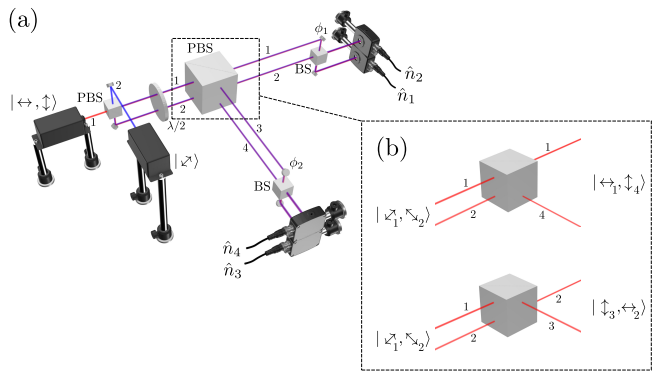}
    \caption{
		     \textbf{(a)} Conceptual sketch of the interferometric setup. 
		     In the first measurement step, a diagonally-polarized attenuated laser beam is incident on 
				 input-port 2 (blue beam) and split 50:50 on the first PBS. The $\lambda/2$-waveplate 
				 brings the separate beams 1 and 2 back into diagonal polarization and they are again 
				 split with a 50:50 ratio on the second PBS. The beams acquire a phase-difference $\phi_1$ ($\phi_2$) 
				 between the paths 1 and 2 (3 and 4) and after recombination on the non-polarizing beam 
				 splitters we observe typical Mach-Zehnder interference fringes in the differences of 
				 the single-photon counts. In the second stage, two indistinguishable photons in 
				 orthogonal polarization are injected simultaneously into input-port 1 (red beam). 
				 Accordingly, they are separated at the first PBS and rotated to (anti-)diagonal 
				 polarization at the $\lambda/2$-waveplate. At the second PBS we consider two possible 
				 paths of the two-photon state. \textbf{(b)} One path where the photon in 
				 beam 1 is transmitted and the photon in beam 2 is reflected and the other path, where 
				 the photon in beam 1 is reflected and the photon in beam 2 is transmitted. Both 
				 possible paths contribute to the coincidence rates between detectors in the (1,2)-arm 
				 and the (3,4)-arm. The interferometric superposition of these two paths, ultimately 
				 reveals the exchange phase of photons.}
    \label{fig:setup_simple}
\end{figure}

\begin{figure*}
    \centering
    \includegraphics[width=\linewidth]{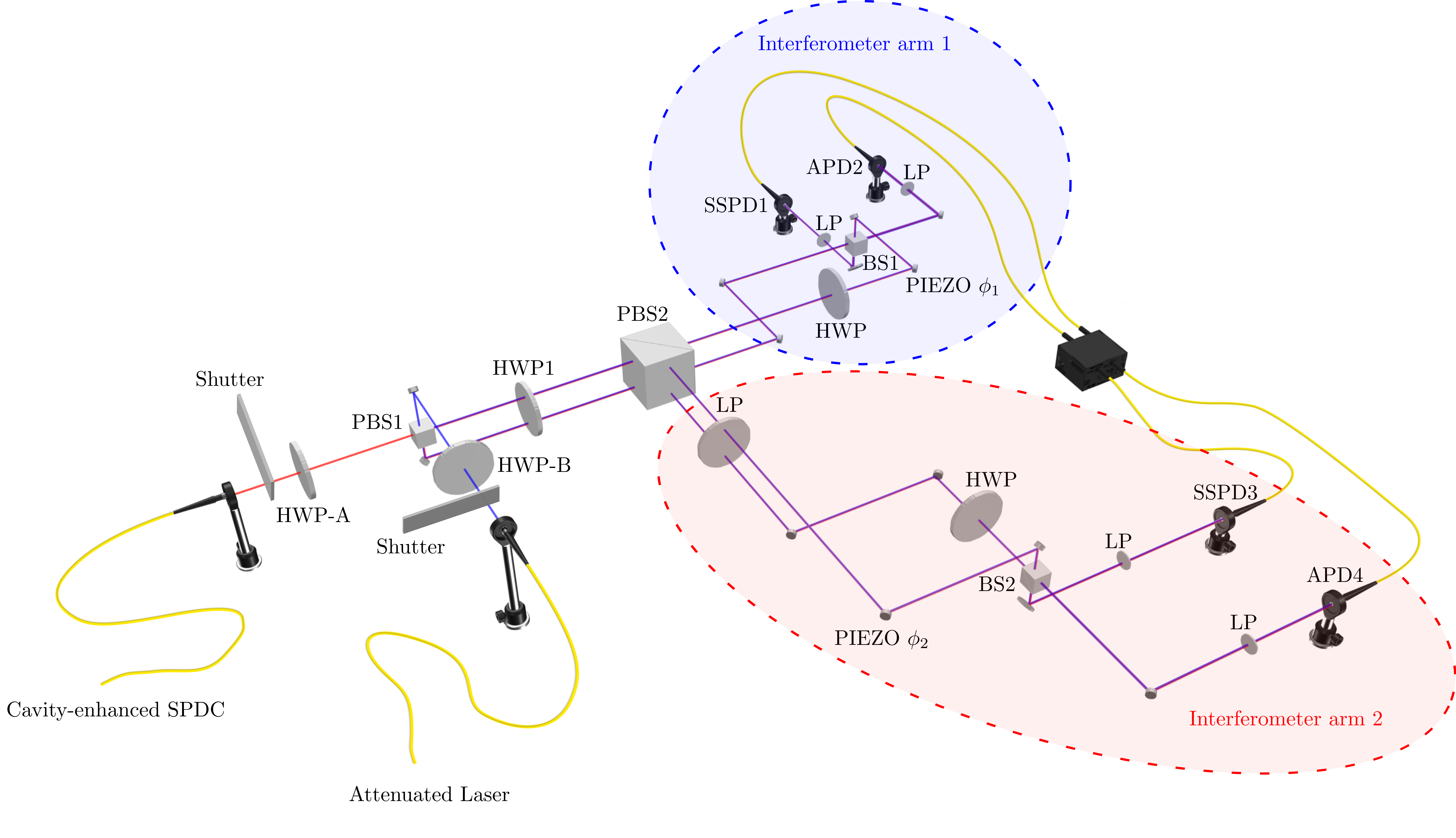}
    \caption{Complete interferometer setup. Indistinguishable photon pairs (red line) are launched into the 
		     interferometer (post-selected pair detection rate of $\approx 4200 ~\text{pairs}/\text{min}$) using a polarization maintaining single mode fiber. 
		      The half-wave-plate-A (HWP-A) rotates the polarization of the photons to horizontal/vertical
				 so that PBS1 splits them deterministically into two paths, 
				 as required by state-dependent transport protocol. A linear polarizer in the interferometer 
				 arm 2 is used to suppress the photons reflected at PBS2 with the undesired polarization. 
				 Both interferometer arms contain a mirror attached to a piezo-element to enable the 
				 deterministic control of the reference phases $\phi_1$ and $\phi_2$. A periscope in 
				 each arm combines the different heights. Due to the orientation of the mirrors in the 
				 periscopes, the polarization of photons in the upper path changes with respect to photons 
				 in the lower path. This is compensated by an additional HWP so that the two paths can 
				 interfere at the beam splitters BS1 and BS2. Linear polarizers (LP) after each output 
				 of BS1 and BS2 filter out photons with an undesired polarization. The photons are collected 
				 into single mode fibers -- which also facilitates the optimal alignment of the optical 
				 paths -- and guided to the individual detectors. Each arm is monitored by one avalanche photodiode 
				 (APD) and one superconducting single photon detector (SSPD), in order to maintain the 
				 symmetry between the interferometer arms. We correlate the detector signals with a time 
				 tagging module (PicoQuant MultiHarp 150). The second input port (blue line) of the 
				 interferometer connects the attenuated laser beam with the interferometer. HWP-B rotates 
				 the polarization of the beam to diagonal polarization, leading to an equal splitting 
				 ratio at PBS1. Finally, automated mechanical shutters at the input ports control whether the photon pairs or 
				 the attenuated laser are sent into the interferometer.}
    \label{fig:setup_complete}
\end{figure*}

Our experimental setup is based on the state-dependent-transport protocol as laid out in \cite{CFRoos}. 
It consists of two coupled Mach-Zehnder interferometers sharing the input ports and coupled to 
each other by a swap beam splitter (swap-BS), which is realized using a polarization beamsplitter (PBS), 
Fig.~(\ref{fig:setup_simple}-a). 
Here, we use the convention that the PBS's reflect (transmit) vertically-polarized (horizontally-polarized) 
light. 
As input we launch two collinear orthogonally-polarized photons into port 1, which are routed 
to paths $1$ and $2$ using a PBS, yielding the state $\ket{\leftrightarrow_1,\updownarrow_2}$. 
Using a half-wave plate (HWP), we transform this latter state into $\ket{\neswarrow_1,\nwsearrow_2}$, 
which is a diagonally-anti-diagonally polarized two photon state on paths 1 and 2, respectively. 
Crucially, impinging the state $\ket{\neswarrow_1,\nwsearrow_2}$ onto the swap-BS generates the 
reference state $\ket{\leftrightarrow_1,\updownarrow_4}$ and a permuted version of it 
$\ket{\updownarrow_3,\leftrightarrow_2}$, Fig.~(\ref{fig:setup_simple}-b).
Notice we disregarded the contributions $\ket{\leftrightarrow_1,\leftrightarrow_2}$ and $\ket{\updownarrow_3,\updownarrow_4}$, where both photons are transmitted or reflected on the swap-BS, in post-selection.\\
As required by the state-dependent-transport protocol, up to this point the two-particle states 
have not overlapped. Consequently, the total wave function lacks any meaningful symmetry property 
\cite{Mirman}. We then use two 50-50 beamsplitters to overlap the reference and the permuted 
states. To obtain the relative phase we need a convenient observable for the interfered states 
such that its expectation value yields a cross-term whose argument is the total phase. This 
observable is found to be the combined photon coincidence rate between the four detectors at 
the outputs
\begin{align}
\begin{split}
    \braket{\hat{\Pi}}&\equiv\langle \hat{n}_1\hat{n}_4+\hat{n}_2\hat{n}_3-\hat{n}_1\hat{n}_3-\hat{n}_2\hat{n}_4 \rangle \\
    &= \frac{1}{2}\cos(\phi_1+\phi_2+\pi-\phi_x),
    \end{split}
\label{eq:pair}
\end{align}
where $\phi_1$ and $\phi_2$ are two known reference phases, which are adjusted deterministically 
using the two mirrors in arms 1 and 2 attached to two piezo-elements. In Fig.~(\ref{fig:setup_complete}) we present a detailed description of the complete interferometer implemented for our experiments, including  
the state preparation and the measurement stage.\\

\begin{figure*}
    \centering
    \includegraphics[width=\linewidth]{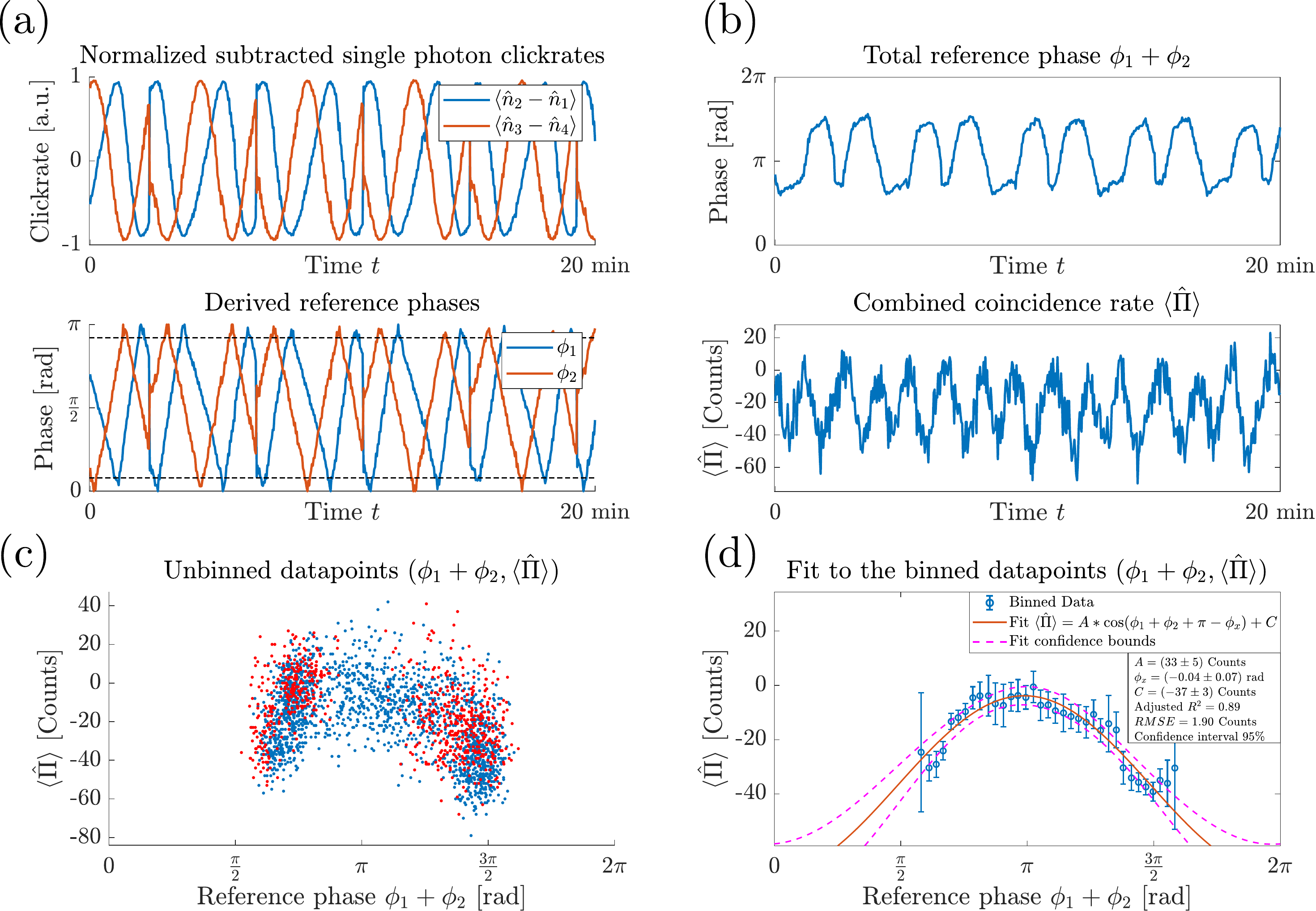}
    \caption{Measurement results. \textbf{(a)} Mach-Zehnder interference fringes observed over a time interval of 20 min, while the attenuated laser is fed into the setup. 
		      On the top we show the evolution of the differences of the normalized single-photon 
					click-rates $\braket{\hat{n}_2-\hat{n}_1}$ (blue curve) and $\braket{\hat{n}_3-\hat{n}_4}$ 
					(red curve) while changing the voltage applied to the piezo-elements. Note, that 
					the discontinuous jumps correspond to the points where the applied saw-tooth voltage 
					signal reverts to its minimum. From these measurements we obtain the corresponding 
					reference phases $\phi_1$ and $\phi_2$ (bottom). 
					\textbf{(b)} Total reference phase (top) and combined coincidence rate $\braket{\hat{\Pi}}$ (bottom) as they evolve in 
					time. 
					\textbf{(c)} Scatter-plot of all measured 
					value pairs $(\phi_1+\phi_2,\braket{\hat{\Pi}})$ ($\approx$ 90 min measurement time in total). 
					Since the slope of $\cos(\phi_{1(2)})$ vanishes at $\phi_{1(2)}=0,\pi$, the associated uncertainty 
					in the estimation of $\phi_1$ and $\phi_2$ diverges at these points. 
					Therefore we consider only data points where $\phi_1$ and $\phi_2$ are within an interval of $[t,\pi-t]$ with $t=0.25$ rad (blue points). 
					This interval is also indicated by the dashed horizontal lines in (a).
					\textbf{(d)} We sort the data points into bins with respect to the total reference phase and 
					perform a least-square error fit of $\braket{\hat{\Pi}}=A\cos(\phi_1+\phi_2+\pi-\phi_x)+C$, which 
					yields $\phi_x=(-0.04 \pm 0.07)$ rad (95\% confidence interval, adjusted $R^2=0.89$, $RMSE=1.9$ Counts) and confirms the 
					symmetry of the two-photon wavefunction.}
    \label{fig:results}
\end{figure*}

The reference phases $\phi_1$ and $\phi_2$ are obtained by separately launching a diagonally-polarized, 
attenuated laser beam into the input-port 2 and taking the difference of the single-photon click-rates 
between the detectors $\left(\hat{n}_1,\hat{n}_2\right)$ and $\left(\hat{n}_3, \hat{n}_4\right)$ 
to give $\langle \hat{n}_2 - \hat{n}_1 \rangle = \frac{1}{2}\cos \left(\phi_1 \right)$ and 
$\langle \hat{n}_3 - \hat{n}_4 \rangle = \frac{1}{2} \cos \left(\phi_2 \right)$, see Fig.~(\ref{fig:results}-a). 
Since the attenuated laser source has a small probability of emitting two photons (an unwanted 
contribution at this measurement stage) we discarded all detection events where two detectors 
clicked within a window of 400 ns.\\ 
In Eq.~(\ref{eq:pair}) we have included an additional phase $\pi$ to account for the geometric phase 
contributed by the physical swap-operation on the two-photon states. Indeed, and as alluded to above, 
the relative phase between the states is not only determined by the particles' fundamental statistics 
$\phi_x$, but also by the dynamic phase $\phi_d$ and the geometric phase $\phi_g$, as defined by 
Aharonov and Anandan \cite{AharonovAnandan}. 
In general, the physical swapping of the quantum states of two indistinguishable particles yields $\phi_g=\pi$, while the 
dynamic phase $\phi_d=0$ vanishes \cite{SI}.\\

To measure $\hat{\Pi}$ as a function of the total reference phase $(\phi_1 +\phi_2)$, we first send 
calibration photons to determine the actual phases $\phi_1$ and $\phi_2$. Next, we launch photon 
pairs (post-selected detection rate of $\approx 4200 ~\text{pairs}/\text{min}$) and measure photon coincidences to obtain $\braket{\hat{\Pi}}$. By repeating this process 
and changing the voltages applied to the two piezo elements after each measurement, we collect the coincidences for several 
values of $(\phi_1 +\phi_2)$, as shown in Fig.~(\ref{fig:results}-b). 
In order to exclude indeterministic thermal fluctuations of $\phi_1$ and $\phi_2$ we set the accumulation time 
for one measurement point to $\approx$ 1 seconds, which is much smaller than the time-scale (several 
hours to days) over which thermal phase drifts have been observed in our setup. 

Notably, for the observable $\hat{\Pi}$ the losses, dark counts, and imperfect indistinguishability 
(86 $\pm$ 5 \% in our case, where 100\% corresponds to perfect indistinguishability) only contribute as a visibility reduction, and as an offset in the 
$\braket{\hat{\Pi}}$-signal, which is independent of $\phi_1$ and $\phi_2$. Since the particle 
exchange phase $\phi_x$ is given as a horizontal displacement of $\braket{\hat{\Pi}}$ along the 
$(\phi_1+\phi_2)$-axis, our results are therefore robust against systematic errors due to 
experimental imperfections \cite{SI}.\\
Fig.~(\ref{fig:results}-c) depicts the measured coincidence rate, $\braket{\hat{\Pi}}$, for 
different values of $(\phi_1 +\phi_2)$ (blue dots). The data points are sorted into bins
of width $0.1$ radians, with respect to the sum $(\phi_1 +\phi_2)$, and the mean value of the bins is calculated with the standard 
error as an uncertainty, Fig.~(\ref{fig:results}-d). To model the measured data and to
determine the phase experimentally we use a slight variation of Eq.~(\ref{eq:pair}), 
$\braket{\hat{\Pi}}=A \cos(\phi_1+\phi_2-\phi_{x}+\pi)+C$, where the amplitude A has to be 
positive and $C$ describes a constant offset. Both constants characterize the combined effects of 
the brightness of the two-photon source, detection efficiencies, detector noise, distinguishability 
of the photon pairs and the integration times. As explained in \cite{SI} these effects do 
not contribute to a horizontal displacement of the signal (along the $(\phi_1+\phi_2)$-axis) 
and, therefore, have no systematic impact on the obtained value for the EP $\phi_x$. This feature makes the present interferometric method robust against experimental imperfection. 
The fit 
(red solid line) in Fig.~(\ref{fig:results}-d) reveals an exchange phase of $\phi_{x}=(-0.04 \pm 0.07)$ radians (95\% confidence 
interval). This result includes an exchange phase of zero revealing the symmetric nature of 
the two-photon state and agrees with the expectation that photons are indeed bosons. 
Furthermore, this result demonstrates that it is crucial to consider the geometric phase of 
the swapping process, otherwise our measurements would lead to the erroneous conclusion that 
two-photon states are anti-symmetric. \\
It is interesting to note that many textbooks introduce the symmetrization postulate stating that quantum mechanical systems comprising $N$ identical particles are either totally symmetric or anti-symmetric under the exchange of any pair of particles \cite{ModernQuantumMechanics,Messiah}. Such statement seems to imply that the physical situation must remain unaffected if the particles are physically exchanged. However, as demonstrated here, that is not the case, and our work will serve as reference to correctly address the symmetrization postulate.
Furthermore, our results provide a first bound for a possibly non-vanishing exchange phase of photons, and is a starting point for precision measurements on tests of the symmetry of multi-particle wavefunctions. \par 

We have developed an interferometric technique to directly measure the particle exchange phase 
of photons. To the best of our knowledge, this is the first direct interferometric measurement 
of the particle exchange phase. Within the margin of error our results confirm the symmetric 
nature of states that consist of two indistinguishable photons. Additionally, we demonstrated 
that it is crucial to consider the additional geometric and dynamic phase accumulated during 
the state-dependent transport protocol. 
Our implementation did not lead to the observation of any deviations from the expected exchange phase. Looking forward, our optical setup may be further improved and optimized towards enhanced accuracy. For example, a brighter and more stable photon pair source would allow for longer measurement times. Also the setup may be implemented using integrated and passively stable optical elements. In the next years a steady increase in accuracy and reduction of the bound for a non-vanishing exchange phase can be achieved. Experimental tests of the exchange phase with other (Fermionic) particles would be highly attractive as well. Finally, our experiment is another example of how non-classical photon pair sources have now entered the field of precision measurements to investigate the validity of very fundamental laws of quantum mechanics \cite{Sinha418}.

\section*{Acknowledgments}
The images of the experimental setup were created with the 3DOptix optical design tool. 
We thank the 3DOptix-Team, who kindly allowed the use of these images in this article.
The authors thank PicoQuant GmbH for providing the MultiHarp 150. \textbf{Funding} C.M. T.K. and O.B. acknowledge support by the German Research Foundation (DFG) Collaborative Research Center (CRC) SFB 787 project C2 and the German Federal Ministry of Education and Research (BMBF) with the project Q.Link.X. 
\textbf{Author contributions:} A.P.L., K.T., O.B. and K.B. initiated the study and guided the work.
K.T., C.M., T.K., M.S. and J.W. designed the interferometer. M.S., C.M.
and T.K. set up the interferometer, C.M. and M.S. performed the optical
measurements. C.M. and K.T. analyzed and interpreted the experimental
data. K.T. and A.P.L. developed the theory. K.T., C.M. and A.P.L. wrote
the manuscript with input from all coauthors.
\textbf{Competing interests:} None declared. 
\textbf{Data and materials availability:} All data needed to evaluate the conclusions in this paper are available in the manuscript and in the supplementary materials.

\appendix

\newpage

\begin{widetext}

\section{Interferometer transformation}
Here, we show that the difference of the single-photon click-rates between the detectors $\hat{n}_2$ and $\hat{n}_1$ (and $\hat{n}_3$ and $\hat{n}_4$) in the interferometer shown in Fig.~(\ref{fig:sketch}) of the main text is given by
\begin{equation}
\langle \hat{n}_2 - \hat{n}_1 \rangle = \frac{1}{2}\cos \left(\phi_1 \right) \text{ and } \langle \hat{n}_3 - \hat{n}_4 \rangle = \frac{1}{2} \cos \left(\phi_2 \right).
\label{eq:single}
\end{equation}
Further we show that
\begin{equation}
    \braket{\hat{\Pi}}\equiv\langle \hat{n}_1\hat{n}_4+\hat{n}_2\hat{n}_3-\hat{n}_1\hat{n}_3-\hat{n}_2\hat{n}_4 \rangle = \frac{1}{2}\cos(\phi_1+\phi_2+\pi-\phi_x).
\end{equation}
To do so, we first derive the input-output relations for single-excitations. We 
denote the different modes in the interferometer by creation operators 
$\hat{a}^\dagger_{x,p}$, where the first index denotes the beam $x=\{1,2,3,4\}$ 
and the second index the polarization $p=\{H,V\}$. It is clear that there are 
4 input-modes: 
$\{ \hat{a}^\dagger_{1,H}, \hat{a}^\dagger_{1,V}, \hat{a}^\dagger_{2,H}, \hat{a}^\dagger_{2,V} \}$ 
and 4 output-modes: 
$\{ \hat{a}^\dagger_{1,H}, \hat{a}^\dagger_{2,H}, \hat{a}^\dagger_{3,V}, \hat{a}^\dagger_{4,V} \}$. 
When optimally aligned, the individual optical devices perform the following 
mode-transformations
\begin{itemize}
    \item PBS: $\hat{a}^\dagger_{x,H} \rightarrow \hat{a}^\dagger_{x,H}$, $\hat{a}^\dagger_{x,V} \rightarrow i \hat{a}^\dagger_{y,V}$
    \item Mirror: $\hat{a}^\dagger_{x,p}\rightarrow i \hat{a}^\dagger_{x,p}$
    \item $\lambda/2$-waveplate: $\hat{a}^\dagger_{x,H} \rightarrow \frac{1}{\sqrt{2}}\left(\hat{a}^\dagger_{x,H}+\hat{a}^\dagger_{x,V} \right)$, $\hat{a}^\dagger_{x,V} \rightarrow \frac{1}{\sqrt{2}}\left(\hat{a}^\dagger_{x,V}-\hat{a}^\dagger_{x,H} \right)$
    \item Phase shifter $\phi_1$: $\hat{a}^\dagger_{x,p} \rightarrow e^{i\phi_1} \hat{a}^\dagger_{x,p}$
    \item Phase shifter $\phi_2$: $\hat{a}^\dagger_{x,p} \rightarrow e^{i\phi_2} \hat{a}^\dagger_{x,p}$
    \item BS: $\hat{a}^\dagger_{x,p} \rightarrow \frac{1}{\sqrt{2}}\left(\hat{a}^\dagger_{x,p}+i \hat{a}^\dagger_{y,p} \right)$
\end{itemize}
\begin{figure}[h!]
    \centering
    \includegraphics[width=\linewidth]{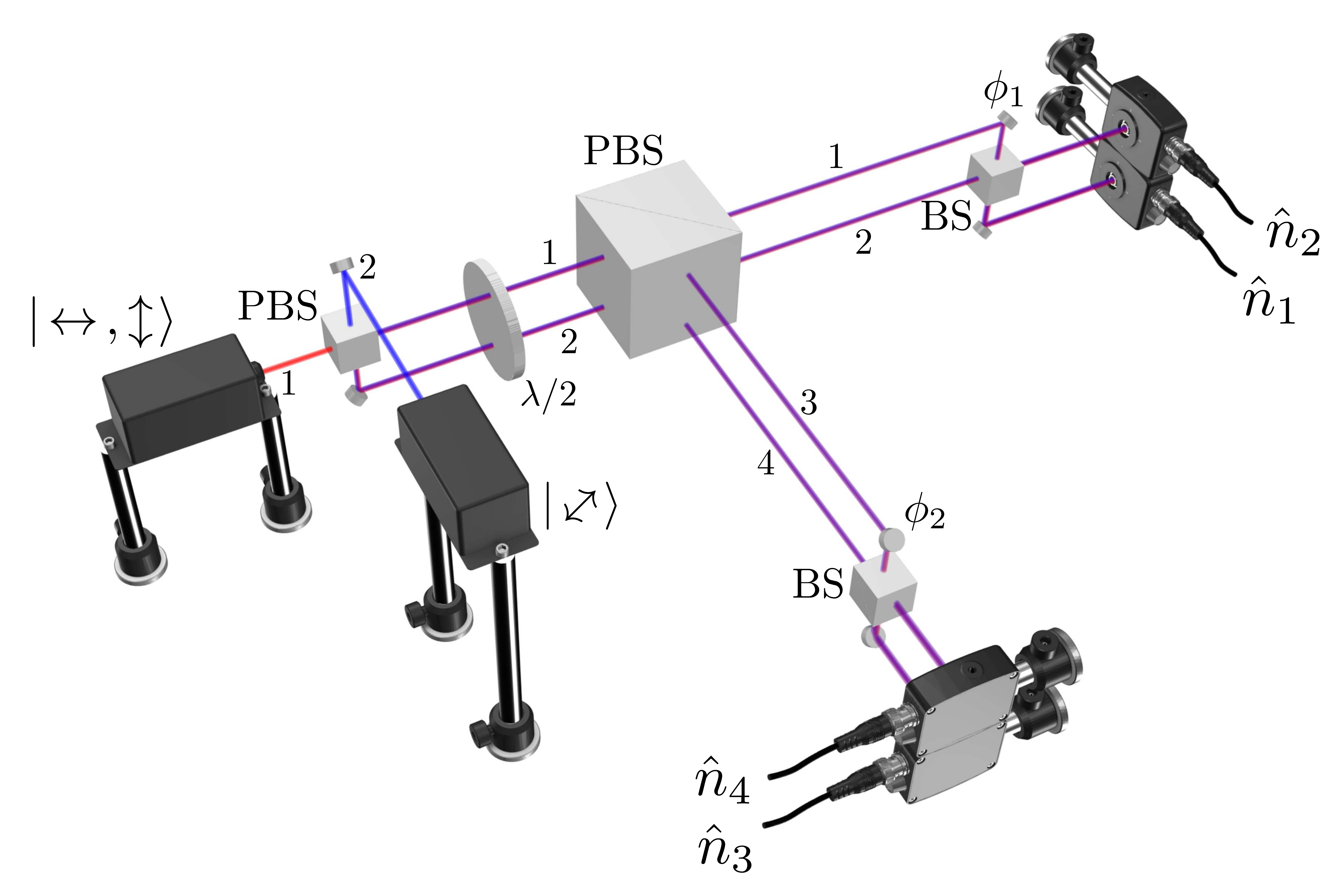}
    \caption{Sketch of the interferometer setup.}
    \label{fig:sketch}
\end{figure}	
Using these relations we derive the complete set of input-output transformations
\begin{align}
    \begin{split}
        \hat{a}^\dagger_{1,H} &\rightarrow \frac{1}{2} \left(i e^{i \phi_1} \hat{a}^\dagger_{1,H} - e^{i\phi_1} \hat{a}^\dagger_{2,H}-\hat{a}^\dagger_{3,V} - i \hat{a}^\dagger_{4,V} \right),\\
        \hat{a}^\dagger_{1,V} &\rightarrow \frac{1}{2} \left( i \hat{a}^\dagger_{1,H} + \hat{a}^\dagger_{2,H}+e^{i \phi_2}\hat{a}^\dagger_{3,V} - i e^{i\phi_2} \hat{a}^\dagger_{4,V} \right),\\
        \hat{a}^\dagger_{2,H} &\rightarrow \frac{i}{2} \left(i \hat{a}^\dagger_{1,H} +\hat{a}^\dagger_{2,H}- e^{i\phi_2}\hat{a}^\dagger_{3,V} +i e^{i\phi_2}\hat{a}^\dagger_{4,V} \right),\\
        \hat{a}^\dagger_{2,V} &\rightarrow \frac{i}{2} \left(-i e^{i \phi_1} \hat{a}^\dagger_{1,H} + e^{i\phi_1} \hat{a}^\dagger_{2,H}- \hat{a}^\dagger_{3,V} -i \hat{a}^\dagger_{4,V} \right).
    \end{split}
    \label{eq:transformation}
\end{align}
In the first measurement stage the input-state is defined by
\begin{equation}
    |\psi_{\text{in}}^{(1)} \rangle = \frac{1}{\sqrt{2}}\left(\hat{a}^\dagger_{2,H} + \hat{a}^\dagger_{2,V} \right) |0\rangle,
\end{equation}
where $|0\rangle$ is the vacuum state. Using Eqs.~(\ref{eq:transformation}) we find
%
\begin{equation}
    | \psi_{\text{out}}^{(1)} \rangle = \frac{i}{2 \sqrt{2}} \left(i(1-e^{i \phi_1})\hat{a}^\dagger_{1,H} + (e^{i \phi_1}+1)\hat{a}^\dagger_{2,H}- (e^{i \phi_2}+1)\hat{a}^\dagger_{3,V}+ i(e^{i \phi_2}-1)\hat{a}^\dagger_{4,V} \right) | 0\rangle.
    \label{eq:outputstate-single}
\end{equation}
%
The single-photon click-rates 
$\langle \hat{n}_i \rangle=\langle \psi_{\text{out}}^{(1)}|\hat{n}_i | \psi_{\text{out}}^{(1)}\rangle$ 
at the four detectors $i=1,2,3,4$ are then 
\begin{align}
\begin{split}
    \langle \hat{n}_1 \rangle = \frac{1}{4}\left(1- \cos \phi_1 \right), & \qquad \langle \hat{n}_2 \rangle = \frac{1}{4}\left(1+ \cos \phi_1 \right),\\
    \langle \hat{n}_3 \rangle = \frac{1}{4}\left(1+ \cos \phi_2 \right), & \qquad   \langle \hat{n}_4 \rangle = \frac{1}{4}\left(1- \cos \phi_2 \right),
\end{split}
\end{align}
and combining these expressions we obtain Eqs.~(\ref{eq:single}). \par
For the second measurement stage, the interferometer is excited by the two-photon input-state
\begin{equation}
    | \psi_{\text{in}}^{(2)} \rangle = \hat{a}^\dagger_{1,H} \hat{a}^\dagger_{1,V} | 0 \rangle. 
    \label{eq:twophoton-input}
\end{equation}
By using the input-output transformations given in Eq. (\ref{eq:transformation}) we obtain 16 terms. 
However, since we are only interested in terms corresponding to the coincidences 
$\hat{n}_1 \hat{n}_3$, $\hat{n}_1 \hat{n}_4$, $\hat{n}_2 \hat{n}_3$ and $\hat{n}_2 \hat{n}_4$, 
we ignore half of the terms to obtain
%
\begin{align}
\begin{split}
|\psi_{\text{out}}^{(2)}\rangle =& \frac{1}{4}\Big( i e^{i(\phi_1+\phi_2)} \hat{a}^\dagger_{1,H}\hat{a}^\dagger_{3,V}+ e^{i(\phi_1+\phi_2)} \hat{a}^\dagger_{1,H}\hat{a}^\dagger_{4,V} -e^{i(\phi_1+\phi_2)} \hat{a}^\dagger_{2,H}\hat{a}^\dagger_{3,V} +i e^{i(\phi_1+\phi_2)} \hat{a}^\dagger_{2,H}\hat{a}^\dagger_{4,V} \\
&-i \hat{a}^\dagger_{3,V}\hat{a}^\dagger_{1,H} -\hat{a}^\dagger_{3,V}\hat{a}^\dagger_{2,H} + \hat{a}^\dagger_{4,V}\hat{a}^\dagger_{1,H} -i \hat{a}^\dagger_{4,V}\hat{a}^\dagger_{2,H} \Big) |0\rangle.
\end{split}
\end{align}
%
Using the definition of the exchange phase $\hat{a}_1^\dagger \hat{a}_2^\dagger = e^{i\phi_x} \hat{a}_2^\dagger \hat{a}^\dagger_1$ we reduce the number of terms to four
%
\begin{align}
\begin{split}
    |\psi_{\text{out}}^{(2)}\rangle =& \frac{1}{4} \Big(i (e^{i(\phi_1+\phi_2)}-e^{i\phi_x}) \hat{a}^\dagger_{1,H}\hat{a}^\dagger_{3,V}+ (e^{i(\phi_1+\phi_2)}+e^{i\phi_x}) \hat{a}^\dagger_{1,H}\hat{a}^\dagger_{4,V} \\ &-(e^{i(\phi_1+\phi_2)}+e^{i\phi_x}) \hat{a}^\dagger_{2,H}\hat{a}^\dagger_{3,V} +i (e^{i(\phi_1+\phi_2)}-e^{i\phi_x}) \hat{a}^\dagger_{2,H}\hat{a}^\dagger_{4,V} \Big)|0\rangle.
\end{split}
\end{align}
%
Thus we have
%
\begin{align}
    \begin{split}
        \langle \hat{n}_1 \hat{n}_3 \rangle = \frac{1}{8}\left(1- \cos( \phi_1+\phi_2-\phi_x )\right) & \qquad \langle \hat{n}_1 \hat{n}_4 \rangle = \frac{1}{8}\left(1+ \cos( \phi_1+\phi_2-\phi_x)  \right)\\
    \langle \hat{n}_2 \hat{n}_3 \rangle = \frac{1}{8}\left(1+ \cos( \phi_1+\phi_2-\phi_x) \right) & \qquad  \langle \hat{n}_2  \hat{n}_4 \rangle = \frac{1}{8}\left(1- \cos (\phi_1+\phi_2-\phi_x) \right)
    \end{split}.
\end{align}
%
Combining these expressions yields
\begin{equation}
\braket{\hat{\Pi}}\equiv \braket{\hat{n}_1\hat{n}_4 + \hat{n}_2 \hat{n}_3 - \hat{n}_1\hat{n}_3 - \hat{n}_2\hat{n}_4} = \frac{1}{2} \cos(\phi_1+\phi_2-\phi_x).
\end{equation}
Note, that in this result we have not included the geometric phase arising from the SWAP operation. 


\section{Single photon measurement with losses and dark counts} 
In what follows, we model the interferometer considering losses and dark counts. 
In principle every device in the setup contributes individually to losses and dark 
counts - either by detecting environmental photons or by 
deflecting photons out of the interferometer. Similarly, misalignment
in the polarization, deviations from a perfect 50:50 splitting ratio at the beam 
splitters and imperfect beam overlap contribute as effective losses and dark counts at the detectors. A detailed and 
comprehensive analysis of every individual component is not feasible and furthermore 
is not guaranteed to properly characterize the interferometer at every point in time 
during the measurement. Thus, for each detector $\hat{n}_i$, we define an effective 
quantum efficiency $\eta_i$ and an effective dark count rate $\nu_i$ \cite{Sperling2013}. We then show 
that the time-averaged click-rates $\bar{c}_i$ and visibilities $v_i$ of the interference 
fringes in the detectors only depend on $\eta_i$, $\nu_i$ and the average photon number of the attenuated laser beam $|\beta|^2$. As such, the parameters 
$(\bar{c}_i,v_i)$ are phenomenological fit-parameters which completely characterize 
the imperfections of the setup \cite{Omar2019}.\\
In the first measurement stage, we implement the single photon source by an attenuated 
laser \cite{Heilmann}, which we model as a coherent state with average photon number $|\beta|^2\approx0.1$. Correspondingly, the input-state 
is a two-mode coherent state 
\begin{equation}
    | \psi^{(1)}_{\text{in}} \rangle = \hat{D}_{2,H+V}(\beta) |0\rangle = | 0,0,\beta/\sqrt{2},\beta/\sqrt{2}\rangle,
\end{equation}
where we have used the Glauber displacement operators \cite{Glauber1963}

\begin{equation}
\hat{D}_{2,H+V}(\beta)=\exp\left( \frac{\beta}{\sqrt{2}} \left(\hat{a}^\dagger_{2,H} + \hat{a}^\dagger_{2,V} \right)-\frac{\beta^*}{\sqrt{2}} \left(\hat{a}_{2,H} + \hat{a}_{2,V} \right)\right).
\end{equation}
Using the ideal interferometer transformation given in Eqs.~(\ref{eq:transformation}) we find 
the four-mode coherent state 
\begin{equation}
    | \psi^{(1)}_{\text{out}} \rangle = \left|\beta_1,\beta_2,\beta_3,\beta_4 \right\rangle = \left| \frac{-\beta(1-e^{i\phi_1})}{2\sqrt{2}},\frac{i\beta(e^{i\phi_1}+1)}{2\sqrt{2}}, \frac{-i\beta(e^{i\phi_1}+1)}{2\sqrt{2}}, \frac{-\beta(e^{i\phi_1}-1)}{2\sqrt{2}}\right\rangle
    \label{eq:output-coherent}
\end{equation}
at the output, which is in complete analogy to the result for a single-photon in 
Eq.~(\ref{eq:outputstate-single}). Now we are interested in the effective single-photon, 
non-coincidence click-rate, in the $i$'th detector
\begin{equation}
    \hat{c}_i=\left(\eta_i \hat{n}_i + \nu_i \right)  e^{-\sum_{i=1}^4 \eta_j \hat{n}_j +\nu_j},
\end{equation}
where we have considered the effective quantum efficiencies and dark count rates. 
To calculate the expectation value of this operator, we utilize the Glauber-Sudarshan 
$P$-function representation \cite{Glauber1963,Sudarshan}. As such, an arbitrary M-mode coherent state 
$|\vec{\beta}\rangle =|\beta_1,\ldots,\beta_M \rangle$ can be written as
\begin{equation}
    P_{\vec{\beta}}(\vec{\alpha}) = \prod_{j=1}^M \delta^2(\alpha_j-\beta_j),
\end{equation}
where $\delta^2(x)=\delta(x)\delta(x^*)$. Using the optical equivalence theorem \cite{Sudarshan}, 
we can compute the expectation value as
\begin{equation}
    \left\langle: \hat{c}_i :\right\rangle_{\vec{\beta}} = \int P_{\vec{\beta}}(\vec{\alpha}) \left(\eta_i |\alpha_i|^2 +\nu_i \right) e^{-\sum_{j=1}^M \eta_j |\alpha_j|^2 + \nu_j} d^2 \vec{\alpha}.
\end{equation}
The $: \ :$ operator denotes the normal ordering by disregarding the commutation 
relations of the annihilation and creation operators \cite{SubBinomial}. This expression evaluates to
\begin{equation}
\left\langle: \hat{c}_i :\right\rangle_{\vec{\beta}} = \left(\eta_i |\beta_i|^2+\nu_i \right) \prod_{j=1}^M e^{-\eta_j |\beta_j|^2-\nu_j}.
\end{equation}
%
%
%
Explicitly, we find
\begin{align}
    \begin{split}
        \left\langle: \hat{c}_1 :\right\rangle_{\vec{\beta}} = \eta_1 \frac{|\beta|^2}{4} + \nu_1 -\eta_1 \frac{|\beta|^2}{4}\cos(\phi_1)&\qquad \left\langle: \hat{c}_2 :\right\rangle_{\vec{\beta}} = \eta_2 \frac{|\beta|^2}{4} + \nu_2 +\eta_2 \frac{|\beta|^2}{4}\cos(\phi_1) \\
        \left\langle: \hat{c}_3 :\right\rangle_{\vec{\beta}} = \eta_3 \frac{|\beta|^2}{4} + \nu_3 +\eta_3 \frac{|\beta|^2}{4}\cos(\phi_2)&\qquad \left\langle: \hat{c}_4 :\right\rangle_{\vec{\beta}} = \eta_4 \frac{|\beta|^2}{4} + \nu_4 -\eta_4 \frac{|\beta|^2}{4}\cos(\phi_2).
    \end{split}
\end{align}
where we have omitted the common factor $\prod_{j=1}^4 e^{-\eta_j |\beta_j|^2 -\nu_j}$. 
Therefore, the observed interference fringes in the detectors have the visibility
\begin{equation}
    v_i=\eta_i\frac{|\beta|^2}{2} \prod_{j=1}^4 e^{-\eta_j |\beta_j|^2 -\nu_j}
\end{equation}
and time-averaged click-rate of
\begin{equation}
    \bar{c}_i = \left(\eta_i\frac{|\beta|^2}{4}+\nu_i \right)\prod_{j=1}^4 e^{-\eta_j |\beta_j|^2 -\nu_j}.
\end{equation}
As a consequence, we are justified with the phenomenological Ansatz
\begin{align}
\begin{split}
    \langle \hat{n}_1 \rangle= \bar{c}_1-\frac{v_1}{2} \cos(\phi_1)& \qquad \langle \hat{n}_2 \rangle = \bar{c}_2+\frac{v_2}{2} \cos(\phi_1)\\
    \langle \hat{n}_3 \rangle = \bar{c}_3+\frac{v_3}{2} \cos(\phi_2)& \qquad \langle \hat{n}_4 \rangle = \bar{c}_4-\frac{v_4}{2} \cos(\phi_2),\label{eq:counts}
\end{split}
\end{align}
and we obtain the visibilities $v_i$ and time-averaged click-rates $\bar{c}_i$ from 
the measured time-resolved interference fringes. This, in turn, allows us to infer 
the phases $\phi_1$ and $\phi_2$ at a specific point in time via\\
\begin{align}
    \begin{split}
        \phi_1=\arccos \left(2\frac{ \langle \hat{n}_2-\hat{n}_1 \rangle-\bar{c}_2+\bar{c}_1}{v_1+v_2} \right)&\qquad \phi_2=\arccos \left(2\frac{ \langle \hat{n}_3-\hat{n}_4 \rangle-\bar{c}_3+\bar{c}_4}{v_3+v_4} \right)  \label{eq:phis}
    \end{split}
\end{align}\\
\section{Two-photon measurement with losses and dark counts}
As defined in Eq.~(\ref{eq:pair}) the operator $\hat{\Pi}=\hat{n}_1\hat{n}_4+\hat{n}_2\hat{n}_3-\hat{n}_1\hat{n}_3-\hat{n}_2\hat{n}_4$ 
is the observable in an ideal scenario - without any losses or darkcounts. Using the corresponding $P$-function representation, we now derive the expectation 
value of the effective observable
\begin{equation}
    \hat{\Tilde{\Pi}}=\sum_{k,l=1}^4 p_{k,l}  \left(\eta_k \hat{n}_k + \nu_k \right) \left(\eta_l \hat{n}_l + \nu_l \right) e^{-2\sum_{r=1}^4 \eta_r \hat{n}_r + \nu_r},
\end{equation}
with $p_{1,4}=1=p_{2,3}$, $p_{1,3}=-1=p_{2,4}$ and $p_{k,l}=0$ otherwise. Again -- using 
the optical equivalence theorem - we can replace $\hat{\Tilde{\Pi}}$ by the continuous 
function 
\begin{equation}
    \Tilde{\Pi}(\vec{\alpha})=\sum_{k,l=1}^4 p_{k,l}  \left(\eta_k |\alpha_k|^2 + \nu_k \right) \left(\eta_l |\alpha_l|^2 + \nu_l \right) e^{-2\sum_{r=1}^4 \eta_r |\alpha_r|^2 + \nu_r}.
\end{equation}
In contrast to the calculation in the ideal scenario, we now have to consider all 16 terms 
that result from the application of the interferometer transformation Eqs.~(\ref{eq:transformation}) 
to the two-photon input-state Eq.~(\ref{eq:twophoton-input}). In general, we have
\begin{equation}
    |\psi^{(2)}_\text{out} \rangle = \sum_{\mathclap{i=1,j\geq i}}^4 \psi_{ij} \hat{a}^\dagger_i \hat{a}^\dagger_j | 0 \rangle,
\end{equation}
and explicitly
\begin{align}
    \begin{split}
        |\psi_{1,3}|^2 &= |\psi_{2,4}|^2 = \frac{1}{8}(1-\cos(\phi_1+\phi_2-\phi_x)) \\
        |\psi_{1,4}|^2 &= |\psi_{2,3}|^2 = \frac{1}{8}(1+\cos(\phi_1+\phi_2-\phi_x))\\
        |\psi_{1,2}|^2&=|\psi_{3,4}|^2=\frac{1}{8}(1-\cos(\phi_x)) \\
        |\psi_{i,i}|^2&=\frac{1}{16}(1+\cos(\phi_x)), 
    \end{split}
\end{align}
where we have not considered the Aharonov-Anandan geometric phase.
We now translate the output-state $|\psi^{(2)}_\text{out}\rangle$ into the $P$-representation \cite{gerry_knight_2004}:
\begin{equation}
    P(\vec{\alpha})=\sum_{i,j} |\psi_{ij}|^2 e^{|\alpha_i|^2+|\alpha_j|^2} \frac{\partial^2}{\partial \alpha_i \partial \alpha_i^*} \delta(\alpha_i) \delta(\alpha_i^*) \frac{\partial^2}{\partial \alpha_j \partial \alpha_j^*} \delta(\alpha_j)\delta(\alpha_j^*).
\end{equation}
Combining these expressions, we have to evaluate
\begin{equation}
    \langle :\hat{\Tilde{\Pi}}:\rangle = \int  P(\vec{\alpha}) \ \Tilde{\Pi} (\vec{\alpha}) \ \text{d}^2 \vec{\alpha}.
\end{equation}
Using the property of the Dirac-$\delta$ function \cite{gerry_knight_2004}
\begin{equation}
    \int F(\alpha,\alpha^*) \frac{\partial^{2n}}{\partial \alpha^n \partial \alpha^{*n}} \delta(\alpha) \delta (\alpha^*)= \left[ \frac{\partial^{2n}F}{\partial \alpha^n \partial \alpha^{*n} }\right]_{|\alpha,\alpha^* = 0},
\end{equation}
yields
\begin{align}
    \begin{split}
        \langle :\hat{\Tilde{\Pi}}:\rangle &= f_1(\vec{\eta},\vec{\nu}) \cos(\phi_1+\phi_2-\phi_x)+ f_2(\vec{\eta},\vec{\nu}) \cos(\phi_x) +f_3(\vec{\eta},\vec{\nu}), 
    \end{split}
\end{align}
with $f_1(\vec{\eta},\vec{\nu})=\frac{1}{8}e^{-2\sum_i \nu_i} \left(\eta_1+\eta_2+2(\eta_2-\eta_1)(\nu_1-\nu_2)\right)\left(\eta_3+\eta_4+2(\eta_4-\eta_3)(\nu_3-\nu_4) \right)$. The functions $f_2(\vec{\eta},\vec{\nu})$ and $f_3(\vec{\eta},\vec{\nu})$
are given by rather lengthy expressions but they vanish in the case of a symmetric 
interferometer, with $\eta_i=\eta$ and $\nu_i=\nu$
\begin{equation}
    \langle :\hat{\Tilde{\Pi}}:\rangle_{\text{sym}} = \frac{\eta^2 e^{-8 \nu}}{2} \cos (\phi_1 +\phi_2 -\phi_x).
\end{equation}
It is clear, that in this symmetric scenario losses and dark counts only contribute as 
a reduction in the visibility of $\langle : \hat{\Tilde{\Pi}}: \rangle$. But also in the 
asymmetric case, losses and dark counts can only contribute as an additional constant vertical 
off-set. Thus, our measurement of the exchange phase, which is a horizontal off-set, is not 
impacted by systematic errors due to dark counts and losses. Note, that the losses and dark counts in the two-photon 
case are not necessarily equal to those in the single-photon measurement stage, since 
the beam-misalignment at the input-port 1 is most likely different compared to input-port 2. 

\section{Aharonov-Anandan geometric phase of the SWAP-operation}
When physically exchanging two particles, they do not only acquire the particle exchange 
phase $\phi_x=2\pi s$, where $s$ is the particle spin, but also a dynamic phase $\phi_d$ 
and a geometric phase $\phi_g$. While $\phi_x$ is fully determined by the particles' 
statistics (fermionic or bosonic), $\phi_d$ and $\phi_g$ are determined by the physical 
process by which the particles are exchanged, or in other words by the path that the 
particles take around each other. More precisely, the geometric phase $\phi_g$ is determined 
by the shape of the path and the dynamic phase $\phi_d$ by the ``velocity'' along the path. 
We now proceed to show that the dynamic phase of the physical swap-operation of two 
photons vanishes, while the geometric phase yields exactly $\pi$. Note, that these phases 
are acquired independent of the particles' statistics. \\
We start with the quantum SWAP-gate, which can be represented by the Pauli $\hat{\sigma}_x$ matrix
\begin{equation}
\hat{\sigma}_x=\begin{pmatrix}
0 & 1 \\
1 & 0 
\end{pmatrix}.
\end{equation}
The SWAP-gate performs the desired operation $\ket{\phi_1}=\begin{pmatrix}
1\\0
\end{pmatrix} \stackrel{\text{SWAP}}{\longleftrightarrow} \begin{pmatrix}
0\\1
\end{pmatrix} = \ket{\phi_2}$, where a single particle changes from the state 
$\ket{\phi_1}$ to $\ket{\phi_2}$ and vice versa. Physically, this transformation 
has to be achieved as a time-evolution of the state, namely
\begin{equation}
\hat{U}(t)=e^{-it\hat{\sigma}_x} = \begin{pmatrix}
\cos(t) & i \sin(t) \\
i \sin(t) & \cos(t) 
\end{pmatrix},
\end{equation}
and after $t=\frac{\pi}{2}$ we have again the desired transformation
\begin{equation}
\hat{U} (\pi/2) = i\begin{pmatrix}
0 & 1 \\
1 & 0 
\end{pmatrix}.
\end{equation}
Notice that we have obtained an additional global phase factor $i=e^{i\frac{\pi}{2}}$. 

In terms of creation operators, we may also write 
\begin{align}
\hat{a}_1^\dagger &\stackrel{\text{SWAP}}{\longrightarrow} i \hat{a}_2^\dagger, \\
\hat{a}_2^\dagger &\stackrel{\text{SWAP}}{\longrightarrow} i \hat{a}_1^\dagger .
\end{align}
When performing this operation on two non-interacting photons simultaneously we obtain
\begin{equation}
\hat{a}_1^\dagger \hat{a}_2^\dagger \stackrel{\text{SWAP}}{\longrightarrow} (i\hat{a}_2^\dagger) (i\hat{a}_1^\dagger) = e^{i\pi} \hat{a}_2^\dagger \hat{a}_1^\dagger.
\end{equation}
The total phase acquired in this process is therefore $\phi_{\text{SWAP}} = \pi$. 
We now proceed to show, that this phase comes entirely from the geometric phase 
and is thus independent of the velocity of the transition. As long as the SWAP 
transformation is completed, we will have obtained this extra phase. \\
In general, a normalized state evolves as
\begin{equation}
\ket{\tilde{\psi}(t)}= e^{-i\phi(t)} \ket{\psi(t)},
\label{eq:psitilde}
\end{equation}
where $\phi(t)$ is a time-dependent global phase and $\ket{\psi(t)}$ is the solution 
of the Schrödinger equation. We now assume that a cyclic process from $t=0$ 
to $t=T$, where we reach the initial state again 
\begin{equation}
\ket{\tilde{\psi}(T)} = \ket{\tilde{\psi}(0)} \Leftrightarrow \ket{\psi(T)}=e^{i\Phi} \ket{\psi(0)}.
\end{equation}
In this process the state acquired the global phase \cite{AharonovAnandan}
\begin{align}
\Phi &= \underbrace{\int_0^T  \bra{\tilde{\psi}(t)} i \partial_t \ket{\tilde{\psi}(t)}  \text{d}t}_{\phi_g} - \underbrace{\int_0^T \bra{\psi(t)} \hat{H} \ket{\psi(t)} \text{d}t}_{\phi_d},
\end{align}
which consists of the Aharonov-Anandan geometric phase $\phi_g$ and the dynamic phase 
$\phi_d$. In order to show, that the total phase $\pi$ from the two-particle SWAP 
operation is entirely geometric, we show that the dynamic phase vanishes. 
In other words we evaluate 
\begin{equation}
\phi_d=\int_0^T \bra{\psi(t)} \hat{H} \ket{\psi(t)} \text{d}t.
\end{equation}
The Hamiltonian in our case is the $\sigma_x$ operator acting on each photon
\begin{equation}
\hat{H}=\hat{\sigma}_x\otimes \hat{\mathbf{1}} + \hat{\mathbf{1}}\otimes \hat{\sigma}_x = \begin{pmatrix}
0 & 1 & 1 & 0\\ 
1 & 0 & 0 & 1\\ 
1 & 0 & 0 & 1\\ 
0 & 1 & 1 & 0
\end{pmatrix},
\end{equation}
and the initial state is $\ket{\psi(0)}=\frac{1}{\sqrt{2}}\left(\ket{\phi_1}\otimes\ket{\phi_2}+\ket{\phi_2}\otimes\ket{\phi_1}\right)=(0,1,1,0)^T$. Using the time evolution operator we find
\begin{align}
\phi_d&= \int_0^T \bra{\psi(0)} \hat{U}^\dagger(t) \hat{H} \hat{U}(t) \ket{\psi(0)} \text{d}t\\
&= \int_0^T  \sum_{n,m=1}^4 e^{i\lambda_n t} e^{-i\lambda_m t}\braket{\psi(0)|n}   \bra{n} \hat{H} \ket{m} \braket{m|\psi(0)} \text{d}t.
\end{align}
Here $\ket{n}$ is the $n$'th eigenstate of $\hat{H}$ with eigenvalue $\lambda_n$. 
Since the spectrum of $\hat{H}$ is $\lambda_{1,2,3,4}=\left\{-2,0,0,2\right\}$, we only need 
to consider the terms $n=m=1$ and $n=m=4$. Further, we define $c_n=\braket{\psi(0)|n}$ 
and obtain
\begin{align}
\phi_d&=\int_0^T  |c_1|^2 \lambda_1 + |c_4|^2 \lambda_4 \text{d}t.
\end{align}
Explicitly we have $\ket{1}=\frac{1}{2}\left(-1,1,1,-1 \right)^T$ and 
$\ket{4}=\frac{1}{2}\left(1,1,1,1 \right)^T$ and thus $|c_1|^2=|c_4|^2=\frac{1}{2}$. 
And since $\lambda_1=-\lambda_4=-2$ we have the result
\begin{align}
\phi_d&=\int_0^T  \frac{1}{2}\left( 2 -2\right) \text{d}t = 0.
\end{align}
Therefore, the total phase of the SWAP operation is entirely geometric in 
nature.
\newpage
\section{Distinguishable photons}
In this section we discuss the impact of the imperfect indistinguishability of the 
two-photon source on our measurements. To reiterate, in the ideal case, with 
no losses, dark counts and indistinguishable photons, in the input state $
\hat{a}^\dagger_H \hat{a}^\dagger_V\ket{0}$ we expect to measure
\begin{equation}
    \braket{\hat{\Pi}}=\langle \hat{n}_1\hat{n}_4+\hat{n}_2\hat{n}_3-\hat{n}_1\hat{n}_3-\hat{n}_2\hat{n}_4 \rangle = \frac{1}{2}\cos(\phi_1+\phi_2+\pi-\phi_x).
\end{equation}
The partial distinguishability of the photon pairs potentially contaminates the 
measured value of $\braket{\hat{\Pi}}$ with accidental coincidences from 
the incoherent evolution of distinguishable photon pairs, that did not (or only partially) interfere 
with one-another. However, in the absence of losses and dark counts, the probability 
for a single photon, in either $V$- or $H$-polarization at the input 1, to emerge 
in the detectors is exactly 
$\braket{\hat{n}_1}_{H,V}=\ldots=\braket{\hat{n}_4}_{H,V}=\frac{1}{4}$, 
which follows from the interferometer transformation Eq.~\eqref{eq:transformation}. 
The probability of an accidental (or incoherent) coincidence between any two 
detectors 
$\braket{\hat{n}_i}_H \braket{\hat{n}_j}_V$ 
or 
$\braket{\hat{n}_i}_V \braket{\hat{n}_j}_H$ 
is therefore $\frac{1}{16}$. It is clear, that in this way the contribution of the 
partial distinguishability to $\braket{\hat{\Pi}}$ vanishes exactly.
When considering the effective quantum efficiencies $\vec{\eta}$ and dark counts $\vec{\nu}$, as in the previous 
sections, we find for the distinguishability contribution
\begin{equation}
    \braket{\hat{\Pi}_d}_{\vec{\eta},\vec{\nu}} \propto \frac{1}{16}\left(\eta_2 -\eta_1+4\left( \nu_2 -\nu_1\right)\right) \left( \eta_3 -\eta_4+4\left( \nu_3 -\nu_4\right)\right),
\end{equation}
which is a constant vertical offset (and vanishes for a symmetric 
interferometer) independent of the interferometer phases $\phi_1$ and $\phi_2$. 
Therefore, the imperfect indistinguishability of the two-photon source does
not introduce a systematic error in the estimation of $\phi_x$. Nevertheless a 
high indistinguishability is preferable, since the 
visibility of the signal improves the statistical significance of the result.

\section{Cavity-enhanced SPDC source}
A stable source of pairs of indistinguishable photons is required for measuring the 
exchange phase. We use a 2\,cm long periodically poled potassium 
titanyl phosphate (PPKTP) crystal with type-II phase-matching as an SPDC source, 
which is placed in a triply (pump, signal, idler) resonant cavity. 
This cavity-enhanced SPDC source is designed to emit photon pairs with a bandwidth of 
100\,MHz at the absorption wavelength of the Cesium D1 line (894\,nm). The cavity 
parameters are adjusted so that the signal and idler photons are resonant simultaneously, 
leading to an indistinguishability in frequency at the central cavity mode. Our source has a pair generation rate of 16 kHz/mW into this central mode.
A temperature controlled narrow band Fabry-P\'{e}rot filter cavity suppresses the 
side-modes of the cavity-enhanced SPDC source, see Fig.~(\ref{fig:HOM}-a).
The HOM effect can be used to determine the indistinguishability of the photon pair.
Typical HOM experiments send two photons on a BS from two different 
input ports and investigate the bunching behavior depending on a temporal delay 
between the photons. In contrast, we send the pair on the PBS at the same input 
port and vary the input polarization of the photons. The coincidences of the 
photon pairs are observed with two SSPDs after the PBS. The PBS is splitting the 
pair when the signal and idler photons are horizontally and vertically polarized. If the polarization 
is rotated to diagonal and anti-diagonal, the photons are indistinguishable at the PBS. This leads to photon bunching, which 
can be observed by a reduction of the coincidences to a minimum. 
The coincidence measurements (blue dots in Fig.~(\ref{fig:HOM}-b) for several 
polarizations is fitted (sold orange line), yielding a HOM visibility of 86$\pm$5\%.
Higher values of the indistinguishability were measured with this source before 
but are less stable for a long time measurement. 

\begin{figure}[h!]
    \centering
    \includegraphics[width=\linewidth]{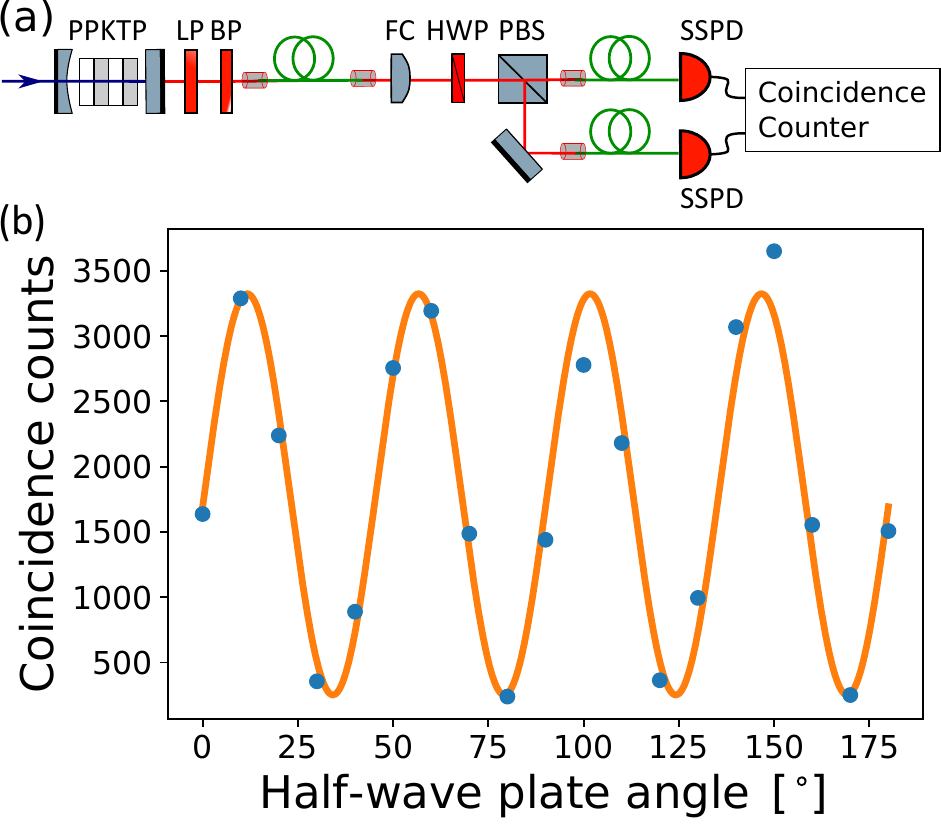}
    \caption{\textbf{(a)} HOM setup to determine the indistinguishability. The SPDC photons 
		are collected by a polarization maintaining single mode fiber and sent through 
		a Fabry-P\'{e}rot filter cavity (FC) with a free spectral range of 28\,GHz 
		and a linewidth of 850\,MHz. The filter cavity is temperature controlled and 
		tuned to the central cavity resonance, where signal and idler are indistinguishable 
		in frequency. The HOM effect can be observed by rotating the HWP and measuring the 
		coincidences after the PBS. The pair is distinguishable when the photons have horizontal and vertical 
		polarization, respectively, and split deterministically at the PBS, leading to the maximum 
		coincidence rate. When the photons have diagonal and anti-diagonal polarization, they are 
		indistinguishable in polarization at the PBS and photon bunching occurs. This leads to a 
		reduction of the coincidence rate to a minimum. \textbf{(b)} The coincidences (blue dots) 
		are measured for several HWP positions. The fluctuation of the coincidences 
		clearly shows the variation of the bunching behavior. A fit (solid orange line) 
		reveals a HOM visibility of 86\,$\pm$\,5\%.}
    \label{fig:HOM}
\end{figure}

\section{Beam Calibration}
We use a strongly attenuated laser beam to characterize the interferometer paths. 
It is important that the calibration beam contains mainly single photons. To purify the 
attenuated laser in post processing, all detection events are discarded when more 
than one photon was detected within 400\,ns. The count rates are detected while 
the phase in each arm is varied by the piezos. We normalize the detected raw counts 
$D$ with the sum of the counts in the corresponding interferometer arm 
\begin{align}
N_i=\frac{D_i}{D_i+D_j},
\end{align} 
where $N_i$ is the normalized count rate, the index $i$ belongs to the considered 
detection path and the index $j$ to the corresponding other detection path of the 
same interferometer arm. This normalization ensures a maximum value of 1 
and renders the result independent of the integration time or slow drifts of the laser 
intensity. Fig.~(\ref{fig:normalizedcounts_measurement}-a) shows the normalized count 
rates for arm 1 for roughly 20 minutes. The discontinuous jumps in the count rates indicate when the 
piezo is moved back to its starting position without additional data being acquired. 
It can be seen clearly that the counts are changing deterministically with the piezo voltage. 
The minimum of $\braket{N_1}$ is reached when $\braket{N_2}$ is at 
its maximum which shows that there is destructive interference at detector~1 when 
there is constructive interference at detector~2. For optimal interference, the 
counts would range from zero to one. However, background and imperfect mode overlap 
at the BS reduces this range. To describe this reduction, we determine the average 
count rate $\bar{c}_i$ and the interference visibility $v_i$ for each detection path with 
the detected counts and Eq.~(\ref{eq:counts}). This was done every 20 minutes 
of the measurement to reduce the influence of slow drifts or power fluctuations of the 
laser. Even though we apply a linear voltage change at the piezos, the piezo response 
is not truly linear. This effect was stronger for the piezo in interferometer arm~1. 
Therefore, we used an average of the ten largest and the ten smallest count rates 
respectively to calculate the parameters $\bar{c}_i$ and $v_i$, instead of using a sinusoidal fit. 

\begin{figure}[h!]
    \centering
    \includegraphics[width=1.0\linewidth]{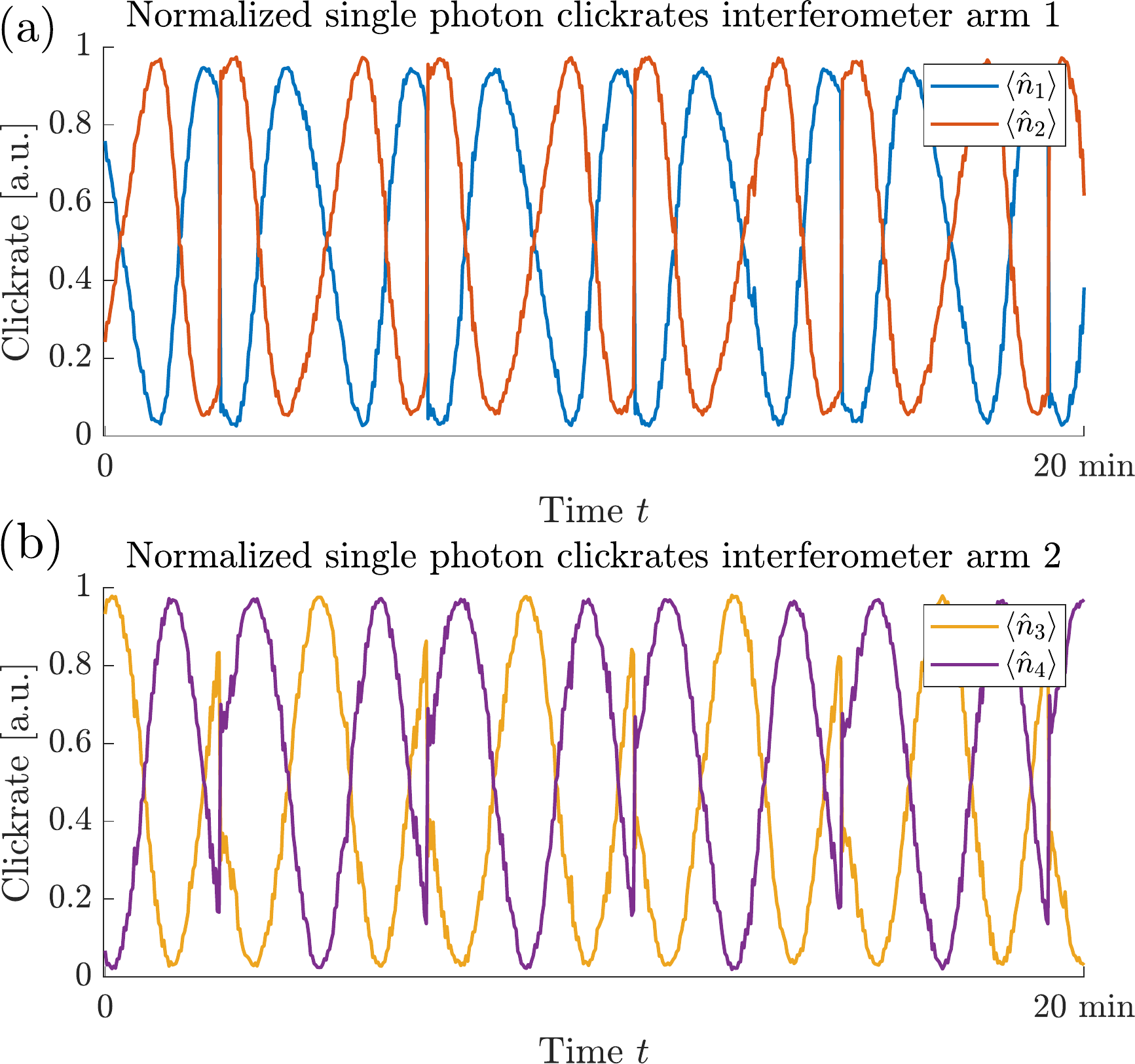}
    \caption{Normalized detected counts for several measurement steps while the piezo 
		voltages are changed after each step. \textbf{(a)} $\braket{N_1}$ (blue line) is at its minimum 
		when $\braket{N_2}$ (red line) reached its maximum and vice versa. This clearly shows the interference 
		of the interferometer arm 1. The discontinuous jumps indicate the returning of the 
		piezos to their starting position without data acquisition. At the beginning of the 
		measurement, the piezo is not at its starting position, since the piezo voltage is 
		controlled independently from the data acquisition. \textbf{(b)} same for interferometer 
		arm 2 with $\braket{N_3}$ (yellow line) and $\braket{N_4}$ (purple line).}
    \label{fig:normalizedcounts_measurement}
\end{figure}

The bottom of Fig.~(3-a) (main text) shows the retrieved reference phases $\phi_1$ and $\phi_2$ from a calibration 
measurement of 20 minutes using Eq.~(\ref{eq:phis}). It can be seen clearly that 
$\phi_1$ is first constantly decreasing until it reaches zero. Afterwards, $\phi_1$ 
increases linearly up to its maximum $\pi$, after which it decreases again. 
$\phi_2$ behaves similar but with an offset compared to $\phi_1$. This phase difference between $\phi_1$ and $\phi_2$ is caused 
by the different relative optical path lengths in the two interferometer arms.

\section{Filter threshold}
The phase retrieval of Eq.~(\ref{eq:phis}) uses an $\arccos$, which exhibits a steep slope at the beginning and the end of its domain. Therefore, small fluctuations during the calibration measurement have a greater influence on the retrieved phase, when the retrieved phase is close to 0 or $\pi$. Hence, we filter out data points when the phases $\phi_1$ and $\phi_2$ are in the range of $0+t$ or $\pi-t$, where $t$ is the filter threshold. The optimal threshold $t$ of such filtering removes the margins with higher uncertainty and simultaneously keeps as much data points as possible to obtain an over-all lower fit uncertainty. Fig.~(\ref{fig:schrittweite0p5}) shows the calculated exchange phase $\phi_x$ for different thresholds~$t$. The lowest uncertainty is reached for a filter threshold of $t=0.25$ rad, with an exchange phase of $\phi_x=(-0.04 \pm 0.07)$ rad. The uncertainty is increasing for $t>0.25$ rad, since the number of data points for calculating $\phi_x$ decreases rapidly. However, all calculated exchange phases $\phi_x$ agree within the limits of errors with the bosonic exchange phase of zero.

\begin{figure}[h!]
	\centering
		\includegraphics[width=1.00\textwidth]{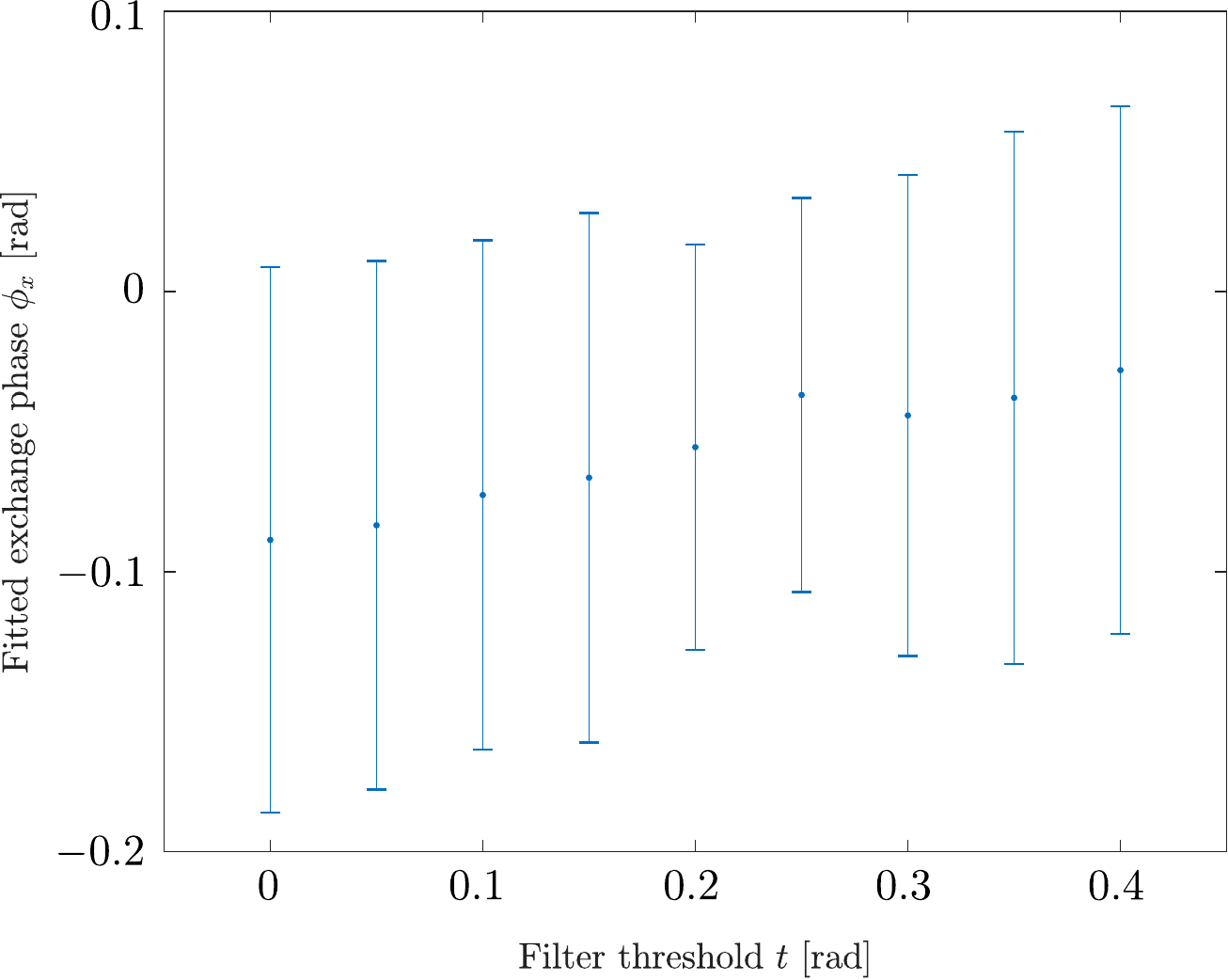}
		\caption{Calculated exchange phase $\phi_x$ for different filter thresholds $t$. The slope of the $\arccos$ diverges for the retrieved reference phases $\phi_{1(2)}$ close to 0 and $\pi$, leading to higher uncertainties for small fluctuations during the calibration measurement. Filtering out these data points can reduce the fit error of the exchange phase $\phi_x$, which is decreasing until $t=0.25$ rad. The uncertainty is increasing again for $t>0.25$ rad since too many data points are filtered out. The exchange phase $\phi_x=(-0.04 \pm 0.07)$ rad at $t=0.25$ rad has the lowest uncertainty and is the main finding of this work.}
	\label{fig:schrittweite0p5}
\end{figure}

\end{widetext}

\bibliographystyle{ieeetr}
\bibliography{exchangephase-arxiv}

\end{document}